\def\ra{\rangle}
\def\la{\langle}
\def\be{\begin{equation}}
\def\ee{\end{equation}}
\def\ba{\begin{array}}
\def\ea{\end{array}}

\def\Cb{{\Bbb C}}

\documentclass[preprint,showpacs,amsmath]{revtex4}
\usepackage{amsfonts}
\usepackage{amssymb}
\def\qed{\leavevmode\unskip\penalty9999 \hbox{}\nobreak\hfill
     \quad\hbox{\leavevmode  \hbox to.77778em{%
               \hfil\vrule   \vbox to.675em%
               {\hrule width.6em\vfil\hrule}\vrule\hfil}}
     \par\vskip3pt}

\input amssym.def
\begin{document}
\title{ Local unitary invariants for multipartite states}
\author{Ting-Gui Zhang$^{1}$}
\author{Ming-Jing Zhao$^{1}$}
\author{Xianqing Li-Jost$^{1}$}
\author{Shao-Ming Fei$^{1,2}$}
\affiliation{$^1$Max-Planck-Institute for Mathematics in the Sciences, 04103 Leipzig, Germany\\
$^2$School of Mathematical Sciences, Capital Normal University,
Beijing 100048, China}

\begin{abstract}
We study the invariants of arbitrary dimensional multipartite quantum states under local unitary transformations. 
For multipartite pure states, we give a set of invariants in terms of singular values of coefficient matrices. For multipartite
mixed states, we propose a set of invariants in terms of the trace of coefficient matrices.
For full ranked mixed states with nondegenerate eigenvalues, this set of invariants is also
the necessary and sufficient conditions for the local unitary equivalence of such two states.
\end{abstract}


\maketitle

\section{Introduction}
Entanglement is one of the most extraordinary features of quantum
theory \cite{EofCon}. The subtle properties of multipartite
entangled states allow for many fascinating applications of quantum
information, such as one-way quantum computing, quantum error
correction and quantum secret sharing \cite{Gott,Raus}. Thus, one of
the main goals in quantum information theory is to gain a better
understanding of the non-local properties of quantum states. According to
the properties of quantum entanglement of multipartite systems,
there are many ways to classify the quantum states, such as local operations and classical
communication (LOCC) and stochastic LOCC (SLOCC) \cite{wdur,F vers,lchen,xiang
Li}.

An important classification of quantum states  is based on the local unitary (LU) transformation. 
That is, given two states $\rho$ and $\rho^\prime$, one asks whether $\rho$
can be transformed into $\rho^\prime$ by LU operations. To solve this problem,
many approaches to construct invariants under local unitary
transformations have been presented in recent years. For example, in \cite{Rains,Grassl} the authors developed 
a method which allows one to
compute all the invariants of local unitary transformations in principle, though
it is not easy to perform operationally. For
multiqubit pure states, the local unitary equivalence problem has been solved in \cite{mqubit}, which is then extended to the arbitrary dimensional case \cite{bliu}. For two qubit mixed states,
a complete set of 18 polynomial invariants is presented in \cite{makhlin}. For high dimensional bipartite mixed states, Zhou \cite{zhou} has
studied the nonlocal properties of
quantum states and solved the local unitary equivalence problem by
presenting a complete set of invariants. Besides, other partial
results have also been obtained for three qubit states \cite{linden},
some generic mixed states \cite{SFG, SFW, SFY} and tripartite
mixed states \cite{SCFW}.
But it is still far away from
understanding the nonlocal properties of multipartite
states completely .

In this article, we study the invariants of arbitrary dimensional multipartite 
quantum states under local unitary operations. For
multipartite pure states, we give a set of invariants in terms of singular values of coefficient matrices. For multipartite
mixed states, we propose a set of invariants in terms of the trace of coefficient matrices, which is also the necessary and sufficient condition of local equivalence 
for full ranked mixed states with nondegenerate eigenvalues.

\section{local unitary invariants for pure state}

First, we consider $n$ partite pure states $|\psi\rangle$ in Hilbert
space $ H_1\otimes H_2\otimes \cdots \otimes H_n$, $|\psi\rangle
=\sum_{s_1=0}^{d_1-1}\sum_{s_2=0}^{d_2-1}\cdots
\sum_{s_n=0}^{d_n-1}a_{s_1s_2\cdots s_n}|s_1s_2\cdots s_n\rangle$,
with $\dim(H_i)=d_i$, and $|s_i\rangle$ the basic vectors of $H_i$, $i=1,\cdots, n$,
$a_{s_1s_2\cdots s_n}\in\Cb$, $\sum_{s_1=0}^{d_1-1}\sum_{s_2=0}^{d_2-1}\cdots
\sum_{s_n=0}^{d_n-1}|a_{s_1s_2\cdots s_n}|^2=1$. Now we associate
$(d_1d_2\cdots d_l)\times (d_{l+1}\cdots d_n)$ coefficient matrix
$M(|\psi\rangle)^{(l)}$ to $|\psi\rangle$ by arranging
$a_{s_1s_2\cdots s_n}$ in lexicographical ascending order, where we
have viewed the indices with respect to the first $l$ qubits as the row ones and the rest indices as
the column ones, $l=1,2,\cdots, [\frac{n}{2}]$. For fixed $l$, all the
possible coefficient matrices can be derived by
$M(|\psi\rangle)^{(l)}$ with permutations \be
{\sigma}={(r_1,c_1)(r_2,c_2)\cdots(r_k,c_k)}\ee where $1\leq r_1 <
r_2 <\cdots<r_k\leq l,\ l<c_1<c_2<\cdots\leq n$, and $(r_i,c_i)$
represents the transposition of $r_i$ and $c_i$. The case
$k=0$ stands for identical permutation, denoted by $\sigma=I$.
Each element in the set $\{\sigma\}$ gives a permutation of $\{1,2,\cdots,n\}$. We
denote $M_{\sigma}(|\psi\rangle)^{(l)}$ the coefficient matrix of
$M(|\psi\rangle)^{(l)}$ under permutation $\sigma$.

For example, for three qubit pure state $|\psi\rangle=\sum_{s_1,s_2,s_3=0}^1 a_{s_1s_2s_3}|s_1s_2s_3\rangle$,
we have
$$
M^{(1)}=\left(
\begin{array}{cccc}
a_{000} & a_{001} & a_{010} & a_{011}\\
a_{100} & a_{101} & a_{110} & a_{111}\\
\end{array}
\right),$$
$$
M^{(1)}_{(1,2)}=\left(
\begin{array}{cccc}
a_{000} & a_{001} & a_{100} & a_{101}\\
a_{010} & a_{011} & a_{110} & a_{111}\\
\end{array}
\right),$$
$$
M^{(1)}_{(1,3)}=\left(
\begin{array}{cccc}
a_{000} & a_{010} & a_{100} & a_{110}\\
a_{001} & a_{011} & a_{101} & a_{111}\\
\end{array}
\right).$$
For four qubit pure state $|\psi\rangle=\sum_{s_1,s_2,s_3,s_4=0}^1 a_{s_1s_2s_3s_4}|s_1s_2s_3s_4\rangle$,
$$M^{(1)}=\left(
\begin{array}{cccccccc}
a_{0000} & a_{0001} & a_{0010} & a_{0011} & a_{0100} & a_{0101} & a_{0110} & a_{0111}\\
a_{1000} & a_{1001} & a_{1010} & a_{1011} & a_{1100} & a_{1101} & a_{1110} & a_{1111}
\end{array}
\right),$$
$$M^{(1)}_{(1,2)}=\left(
\begin{array}{cccccccc}
a_{0000} & a_{0001} & a_{0010} & a_{0011} & a_{1000} & a_{1001} & a_{1010} & a_{1011}\\
a_{0100} & a_{0101} & a_{0110} & a_{0111} & a_{1100} & a_{1101} & a_{1110} & a_{1111}
\end{array}
\right),$$
$$M^{(1)}_{(1,3)}=\left(
\begin{array}{cccccccc}
a_{0000} & a_{0001} & a_{0100} & a_{0101} & a_{1000} & a_{1001} & a_{1100} & a_{1101}\\
a_{0010} & a_{0011} & a_{0110} & a_{0111} & a_{1010} & a_{1011} & a_{1110} & a_{1111}
\end{array}
\right),$$
$$M^{(1)}_{(1,4)}=\left(
\begin{array}{cccccccc}
a_{0000} & a_{0010} & a_{1000} & a_{1010} & a_{0100} & a_{0110} & a_{1100} & a_{1110}\\
a_{0001} & a_{0011} & a_{1001} & a_{1011} & a_{0101} & a_{0111} & a_{1101} & a_{1111}
\end{array}
\right),$$
$$M^{(2)}=\left(
\begin{array}{cccc}
a_{0000} & a_{0001} & a_{0010} & a_{0011}\\
a_{0100} & a_{0101} & a_{0110} & a_{0111}\\
a_{1000} & a_{1001} & a_{1010} & a_{1011}\\
a_{1100} & a_{1101} & a_{1110} & a_{1111}
\end{array}
\right),$$
$$M^{(2)}_{(2,3)}=\left(
\begin{array}{cccc}
a_{0000} & a_{0001} & a_{0100} & a_{0101}\\
a_{0010} & a_{0011} & a_{0110} & a_{0111}\\
a_{1000} & a_{1001} & a_{1100} & a_{1101}\\
a_{1010} & a_{1011} & a_{1110} & a_{1111}
\end{array}
\right),$$
$$M^{(2)}_{(2,4)}=\left(
\begin{array}{cccc}
a_{0000} & a_{0010} & a_{0100} & a_{0110}\\
a_{0001} & a_{0011} & a_{0101} & a_{0111}\\
a_{1000} & a_{1010} & a_{1100} & a_{1110}\\
a_{1001} & a_{1011} & a_{1101} & a_{1111}
\end{array}
\right).$$

If two $n$-partite pure states $|\psi\rangle$ and $|\phi\rangle$ are
LU equivalent, $|\psi\rangle=U_1\otimes U_2\otimes\cdots\otimes
U_n|\phi\rangle$, where $U_1,U_2,\cdots,U_n$ are local unitary
operators in
$SU(d_1,\mathbb{C}),SU(d_2,\mathbb{C}),\cdots,SU(d_n,\mathbb{C})$,
respectively, then the coefficient matrices of $|\psi\rangle$ and
$|\phi\rangle$ satisfy the relation 
\be\label{2}
M_{\sigma}(|\psi\rangle)^{(l)}=U_{\sigma(1)}\otimes U_{\sigma(2)}\otimes\cdots\otimes
U_{\sigma(l)} M_{\sigma}(|\phi\rangle)^{(l)}(U_{\sigma(l+1)}\otimes\cdots \otimes
U_{\sigma(n)})^{T}, \ \ \ \forall \ l
\ee
with superscript $T$ the transpose. From (\ref{2}) we have

(i) rank $M_{\sigma}(|\psi\rangle)^{(l)} =$ rank $M_{\sigma}(|\phi\rangle)^{(l)}$.

(ii)
$Tr[M_{\sigma}(|\psi\rangle)^{(l)}M_{\sigma}(|\psi\rangle)^{(l)\dag}]^{\alpha}=
Tr[M_{\sigma}(|\phi\rangle)^{(l)}M_{\sigma}(|\phi\rangle)^{(l)\dag}]^{\alpha}$,
$\alpha=1,2,\cdots, \min\{d_{\sigma(1)}d_{\sigma(2)}\cdots
d_{\sigma(l)},\ d_{\sigma(1+1)} d_{\sigma(1+2)}\cdots
d_{\sigma(n)}\}$,

(iii) $M_{\sigma}(|\psi\rangle)^{(l)} $ and $M_{\sigma}(|\phi\rangle)^{(l)}$ have the same
singular values, $\forall \ l, \ \sigma$.

The three conditions above are necessary for determining whether two
arbitrary multipartite pure states are local unitary equivalent or
not. In view of the condition (i), if two pure states differ in the
ranks of their corresponding coefficient matrices, 
then they belong to different local unitary equivalent
classes. While from the aspect of (ii), if two coefficient matrices
do not have the same trace relations, they are not local
unitary equivalent. Condition (ii) is strictly stronger than
condition (i) since two matrices with the same rank may have
different trace relations. For example, the three qubit W state $|W\rangle=\frac{1}{\sqrt{3}}(|001\rangle+|010\rangle+|100\rangle)$ and GHZ
state $|GHZ\rangle=\frac{1}{\sqrt{2}}(|000\rangle+|111\rangle)$ have the same rank,
 ${\rm rank} (M_{\sigma}(|W\rangle)^{(l)}) ={\rm rank} (M_{\sigma}(|GHZ\rangle)^{(l)})$,
 but $tr[M_\sigma(|W\rangle)^{(1)}M_\sigma(|W\rangle)^{(1)\dag}]^2=5/9\neq tr[M_\sigma(|GHZ\rangle)^
 {(1)}M_\sigma(|GHZ\rangle)^{(1)\dag}]^2
 =1/2$. Therefore they are not LU equivalent.
In \cite{Linden,SFY}, it has been shown that for bipartite pure states,
condition (ii) is a necessary and sufficient condition of LU equivalece.
Condition (iii) is equivalent to condition (ii) for pure states. 
Nevertheless, both condition (ii)
and (iii) are only necessary for multipartite states. For instance,
consider three qubit pure states
$|\psi_1\rangle=\frac{1}{\sqrt{3}}|000\rangle+\sqrt{\frac{2}{3}}|111\rangle$
and
$|\psi_2\rangle=\frac{1}{\sqrt{3}}(|001\rangle+|010\rangle+|100\rangle)$.
Their coefficient matrices have the same trace relations and the same singular values. But
they can not be transformed into each other neither by LU
operations nor by SLOCC.

\section{local unitary invariants for mixed state}

Now we consider the local unitary invariants for mixed states. Two $n$-partite mixed states $\rho$ and
$\rho^\prime$ in $H_1\otimes H_2\cdots\otimes H_n$ Hilbert space
are said to be equivalent under local unitary transformations if
there exist unitary operators $U_i$ on the i-th Hilbert space such
that 
\be\label{eq} 
\rho^\prime=(U_1\otimes U_2\otimes\cdots\otimes
U_n)\rho(U_1\otimes U_2\otimes\cdots\otimes U_n)^\dag. 
\ee  
One way to deal with the LU equivalence (\ref{eq}) is to use purification. 
After purification, an $n$-partite mixed state becomes an $(n+1)$-partite pure state.
Ref. \cite{zwang} has revealed the relations between the $n$-partite mixed
states and their $(n+1)$-partite purified ones as follows,

\noindent {\bf Lemma 1} If one of the $n$-partite reduced density
matrices of the $(n+1)$-partite pure state $|\psi\rangle$ is local
unitary equivalent to the corresponding $n$ partite reduced density
matrices of the $(n+1)$-partite pure state $|\phi\rangle$, then two
$(n+1)$-partite pure states $|\psi\rangle$ and $|\phi\rangle$ are
also local unitary equivalent.

Employing this Lemma, we have the following result.

\noindent {\bf Theorem 1} An $n$-partite
mixed state $\rho^\prime$ is LU equivalent to $\rho$ if and
only if its purified state is LU
equivalent to that of $\rho$.

\noindent {\bf Proof:} Suppose
$\rho=\sum_{i=1}^{K}\lambda_i|v_i\ra\la v_i|$ and $\rho^\prime
=\sum_{i=1}^{K^\prime}\lambda_i^{\prime }|v_i^\prime\ra\la
v_i^\prime|$ are the spectra decompositions of $\rho$ and
$\rho^\prime$ respectively,
$\sum_{i}\lambda_i=\sum_{i}\lambda_i^{\prime }=1$,
$\lambda_i,\lambda_i^\prime\in\mathbb{R}^+$. Let
$|\psi_0\rangle=\Sigma_{i=1}^K \sqrt{\lambda_i}|v_i\rangle|i\rangle$
be the purification of $\rho$ and
$|\psi_0^\prime\rangle=\Sigma_{i=1}^{K^\prime}
\sqrt{\lambda_i^\prime}|v_i^\prime\rangle|i^\prime\rangle$ the
purification of $\rho^\prime$. If $\rho^\prime$ is LU equivalent to
$\rho$, then by Lemma 1 and the relations
$Tr_{n+1}[|\psi_0\rangle\langle\psi_0|]= \rho$ and
$Tr_{n+1}[|\psi_0^\prime\rangle\langle\psi_0^\prime|]= \rho^\prime$,
we get that $|\psi_0^\prime\rangle$ is LU equivalent to
$|\psi_0\rangle$.

On the other hand, if $|\psi_0^\prime\rangle=U_1\otimes U_2\otimes
U_3\otimes\cdots\otimes U_{n+1}|\psi_0\rangle$, then
$\rho^\prime=Tr_{n+1}[|\psi_0^\prime\rangle\langle\psi_0^\prime|]
=Tr_{n+1}[(U_1\otimes U_2\otimes U_3\otimes\cdots\otimes
U_{n+1})|\psi_0\rangle\langle\psi_0|(U_1\otimes U_2\otimes
U_3\otimes\cdots\otimes U_{n+1})^\dag]
=(U_1\otimes
U_2\otimes\cdots\otimes
U_n)Tr_{n+1}(|\psi_0\rangle\langle\psi_0|)(U_1\otimes
U_2\otimes\cdots\otimes U_n)^{\dag}=(U_1\otimes
U_2\otimes\cdots\otimes U_n)\rho(U_1\otimes U_2\otimes\cdots\otimes
U_n)^{\dag}$. Hence $\rho^\prime$ is LU equivalent to $\rho$.\qed

From Theorem 1 we see that the LU equivalence problem of $n$-partite mixed states
can be transformed into the LU equivalence of $(n+1)$-partite pure states. 
The LU classification for arbitrary dimensional multipartite pure states has been studied
in \cite{bliu} by exploiting the high order singular value decomposition 
technique and local symmetries of the states. Employing the results in \cite{bliu}, 
the LU equivalence problem of mixed states can be solved further.

Besides purification, one may also deal with the LU equivalence of mixed states directly 
in terms of the LU invariants. Next we give a set of invariants in terms of 
the trace relations about the coefficient matrices.

\smallskip
\noindent {\bf Theorem 2} For arbitrary $n$ partite nondegenerate
mixed states $\rho$ with spectra decomposition, $\rho=\sum_{i=1}^{K}\lambda_i|v_i\ra\la v_i|$, $\sum_{i}\lambda_i=1$, $\lambda_i\in\mathbb{R}^+$,
the following quantities are LU invariants,

(a) the rank $K$ of $\rho$;

(b) the eigenvalues $\lambda_i$ of $\rho$, $i=1,\cdots, K$;

(c) $Tr[M_{\sigma}(|v_i\rangle)^{(l)}M_{\sigma}(| v_j\rangle)^{(l)\dag}\cdots
M_{\sigma}(|v_k\rangle)^{(l)}M_{\sigma}(| v_m\rangle)^{(l)\dag}]$, $i,j,\cdots,
k,m=1,\cdots, K$, $\forall \ l,\ \sigma$.

\noindent {\bf Proof}
Let $\rho^\prime=(U_1\otimes U_2\otimes\cdots\otimes U_n)
~\rho~(U_1\otimes U_2\otimes\cdots\otimes U_n)^\dag$, 
where $U_i$, $1\leq i\leq n$, are arbitrary unitary operators. 
Since the eigenvalues of $\rho$ are nondegenerate, so the eigenvalues of $\rho^\prime$ are
$\lambda_i$ with the corresponding eigenvectors $\vert v_i^\prime\rangle=U_1\otimes U_2\otimes\cdots\otimes
U_n\vert v_i\rangle$ up to a global phase, $i=1,\cdots, K$. Equivalently, $M_{\sigma}(|v_i^\prime\rangle)^{(l)}=(U_{\sigma(1)}\otimes U_{\sigma(2)}\otimes \cdots\otimes
U_{\sigma(l)} ) M_{\sigma}(|v_i\rangle)^{(l)} (U_{\sigma(l+1)}\otimes\cdots\otimes U_{\sigma(n)})^T$. Therefore
\be
M_{\sigma}(|v_i^\prime\rangle)^{(l)}
M_{\sigma}(|v_j^\prime\rangle)^{(l)\dag}=U_{\sigma(1)}\otimes U_{\sigma(2)}\otimes
\cdots\otimes U_{\sigma(l)} M_{\sigma}(|v_i\rangle)^{(l)} M_{\sigma}(|v_j\rangle)^{(l)\dag}
(U_{\sigma(1)}\otimes U_{\sigma(2)}\otimes
\cdots\otimes U_{\sigma(l)} )^\dag,\ee
for
$i,j=1,...,K$, which gives rise to $Tr[M_{\sigma}(|v_i^\prime\rangle)^{(l)}
M_{\sigma}(|v_j^\prime\rangle)^{(l)\dag}\cdots M_{\sigma}(|v_k^\prime\rangle)^{(l)}
M_{\sigma}(|v_m^\prime\rangle)^{(l)\dag}]  =Tr[M_{\sigma}(|v_i\rangle)^{(l)}
M_{\sigma}(|v_j\rangle)^{(l)\dag}\cdots M_{\sigma}(|v_k\rangle)^{(l)}
M_{\sigma}(|v_m\rangle)^{(l)\dag}]$. Therefore, the rank, the eigenvalues of $\rho$ 
and the trace of products of the coefficient matrices are invariant under local unitary transformations.\qed

For example, for three qubit mixed states
$\rho_1=\lambda|W\rangle\langle W|+(1-\lambda)|011\rangle\langle
011|$, and $\rho_2=\lambda|GHZ\rangle\langle
GHZ|+(1-\lambda)|011\rangle\langle 011|$, one has,
$tr[M_\sigma(|W\rangle)^{(1)}M_\sigma(|W\rangle)^{(1)\dag}]^2\neq
tr[M_\sigma(|GHZ\rangle)^
 {(1)}M_\sigma(|GHZ\rangle)^{(1)\dag}]^2
 $. Thus they are not LU equivalent. 
 
Generally the invariants in Theorem 2 are only necessary for LU equivalence.
However, for some special sets of multipartite
mixed states, the above invariants are complete.

\noindent {\bf Theorem 3} For two arbitrary $n$ partite nondegenerate and full rank
mixed states $\rho$ and $\rho^\prime$ with spectra decomposition, 
$\rho=\sum_{i=1}^{K}\lambda_i^{\prime }|v_i\ra\la v_i|$, $\rho^\prime=\sum_{i=1}^{K}\lambda_i|v_i^\prime\ra\la v_i^\prime|$, $\sum_{i}\lambda_i=\sum_{i}\lambda_i^{\prime }=1$, $\lambda_i,\lambda_i^\prime\in\mathbb{R}^+$, they are local unitary equivalent if and only if

(a) $\lambda_i=\lambda_i^\prime$, $i=1,\cdots, K$;

(b) $Tr[M_{\sigma}(|v_i\rangle)^{(l)}M_{\sigma}(| v_j\rangle)^{(l)\dag}\cdots
M_{\sigma}(|v_k\rangle)^{(l)}M_{\sigma}(| v_m\rangle)^{(l)\dag}]$, $i,j,\cdots,
k,m=1,\cdots, K$, $\forall \ l, \ \sigma$.

\noindent {\bf Proof}. Here we only need to prove the
sufficiency. If $\rho$ and $\rho^\prime$ satisfy conditions (a) and (b), then they are local unitary equivalent under bipartite partition by Ref. \cite{zhou}. Namely,
\begin{eqnarray}
\rho^\prime=V_{y_1}\otimes V_{y_2} \rho V_{y_1}^\dagger\otimes V_{y_2}^\dagger
\end{eqnarray}
and
\begin{eqnarray}\label{pure eq}
|v_i^\prime\ra=V_{y_1}\otimes V_{y_2} | v_i\rangle, \ \ \forall i,
\end{eqnarray}
for all possible bipartite partitions $(y_1, y_2)$ of the system, where $V_{y_1}$ and $V_{y_2}$ are unitary transformations. Since $\rho$ and $\rho^\prime$ are full ranked, 
$\{| v_i\rangle\}$ and $\{| v_i^\prime\rangle\}$ constitute two orthonormal basis for the whole vector space, which implies that there exists a unique unitary transformation that maps $ |v_i\rangle$ to $| v_i^\prime\rangle$, $\forall i$. The uniqueness of the unitary transformation in Eq. (\ref{pure eq}) makes the whole
unitary transformation a tensor product one acting on the individual subsystems. In this case, $|v_i^\prime\ra=U_1\otimes U_2\otimes \cdots \otimes U_n | v_i\rangle$, $\forall i$, and $\rho^\prime=U_1\otimes U_2\otimes \cdots \otimes U_n \rho U_1^\dagger\otimes U_2^\dagger\otimes \cdots \otimes U_n^\dagger$.
 \qed
\smallskip

\section{Conclusion}

We have investigated the invariants of arbitrary dimensional multipartite quantum states under local unitary operations. We presented the set of coefficient matrices. The singular values of these coefficient matrices
are just the LU invariants for multipartite pure states. For multipartite
mixed states, the trace of the coefficient matrices are LU invariants, which give rise to
the necessary and sufficient conditions for full ranked mixed states with nondegenerate eigenvalues.
As these LU invariants can be explicitly calculated,
our approach gives a simple way in verifying the LU equivalence of given quantum states.


\begin{thebibliography}{99}

\bibitem{EofCon} R. Horodecki, P. Horodecki, M. Horodecki and K. Horodecki, Rev. Mod. Phys. {\bf 81}, 865 (2009).

\bibitem{Gott} D. Gottesman, Ph.D. Thesis, quant-ph/9705052.
\bibitem{Raus} R. Raussendorf and H. J. Briegel, Phys. Rev. Lett.
{\bf 86}, 5188 (2001).
\bibitem{wdur} W. D¨¹r, G. Vidal and J. I. Cirac, Phys. Rev. A {\bf 62}, 062314
(2000).
\bibitem{F vers} F. Verstraete, J. Dehaene, B. DeMoor and H. Verschelde, Phys.
Rev. A {\bf 65}, 052112 (2002).

\bibitem{lchen} L. Chen and Y. X. Chen, Phys.
Rev. A {\bf 74}, 062310 (2006).
\bibitem{xiang Li} X. Li and D. Li, Phys. Rev. Lett. {\bf 108}, 180502
(2012).
\bibitem{Rains} E.M.~Rains, IEEE Transactions on Information Theory, {\bf 46}, 54 (2000).

\bibitem{Grassl} M.~Grassl, M.~R\"otteler and T.~Beth, Phys. Rev. A {\bf 58}, 1833 (1998).

\bibitem{mqubit} B. Kraus, Phys. Rev. Lett. {\bf 104}, 020504 (2010); Phys. Rev. A {\bf 82}, 032121 (2010).

\bibitem{bliu} B. Liu, J.L. Li, X. Li and C.F. Qiao, Phys. Rev. Lett. {\bf 108}, 050501 (2012).

\bibitem{makhlin} Y. Makhlin, Quant. Info. Proc. {\bf 1}, 243 (2002).

\bibitem{zhou} C. Zhou, T.G. Zhang, S.M. Fei, N. Jing and X. Li-Jost, Phys. Rev. A {\bf 86}, 010303 (2012).

\bibitem{linden} N. Linden, S. Popescu and A. Sudbery, Phys. Rev. Lett. {\bf 83}, 243 (1999).

\bibitem{Linden} N. Linden and S. Popescu, Fortsch. Phys. {\bf 46}, 567 (1998).

\bibitem{SFY} S. Albeverio, S.M. Fei, P.Parashar and W.L.Yang, Phys.Rev.A {\bf 68}, 010303 (2003).

\bibitem{SFG} S. Albeverio, S.M. Fei and D.Goswami, Phys.Lett. A {\bf 340}, 37 (2005).

\bibitem{SFW} B.Z. Sun, S.M. Fei, X.Q. Li-Jost and Z.X.Wang, J. Phys. A {\bf 39}, 43 (2006).

\bibitem{SCFW} S. Albeverio, L. Cattaneo, S.M. Fei and X.H. Wang, Int. J. Quant. Inform. {\bf 3}, 603 (2005).

\bibitem{zwang}  Z. Wang, H.P. Wang, Z.X. Wang and S.M. Fei, Chin. Phys. Lett. {\bf 28}, 020302
(2011).


\end{thebibliography}
\end{document}